\documentclass[11pt,a4paper]{article}
\pdfoutput=1
\usepackage{jheppub}
\usepackage{rotating}
\usepackage{array}
\usepackage{amsmath}
\usepackage{gensymb}
\usepackage[normalem]{ulem}
\usepackage{slashed}
\usepackage{booktabs}
\usepackage[pdftex,table,dvipsnames]{xcolor}
\usepackage{units}
\usepackage{multirow}
\usepackage{url}
\usepackage{parskip}
\usepackage[all]{nowidow}
\usepackage{placeins}

\usepackage{xspace}
\usepackage{subfig}
\usepackage{color}

\usepackage{dsfont}
\usepackage{soul}
\usepackage{hyperref}

\allowdisplaybreaks

\def \belletwo {Belle\,II\xspace}
\newcommand{\Br}{\ensuremath{\mathcal{B}}\xspace}

\newcommand\sw[1]{{\textcolor{black}{#1}}}

\preprint{P3H-22-017}

\title{Near or Far Detectors? A Case Study for Long-Lived Particle Searches at Electron-Positron Colliders}

\author[1]{R.~Sch\"afer}
\author[1]{F.~Tillinger}
\author[1,2,3]{S.~Westhoff}

\affiliation[1]{Institute for Theoretical Physics, Heidelberg University, 69120 Heidelberg, Germany}
\affiliation[2]{Institute for Mathematics, Astrophysics and Particle Physics, Radboud University,
Nijmegen, The Netherlands}
\affiliation[3]{\href{https://www.nikhef.nl/en/}{\color{black} Nikhef},
Science Park 105, 1098 XG Amsterdam, The Netherlands\\}

\emailAdd{r.schaefer@thphys.uni-heidelberg.de}
\emailAdd{tillinger@thphys.uni-heidelberg.de}
\emailAdd{susanne.westhoff@ru.nl}

\abstract{We explore the discovery potential for long-lived particles at the 250-GeV ILC. The goal is to investigate possible gains of a dedicated far detector over the main detector ILD. For concreteness, we perform our study for sub-GeV axion-like particles $a$ produced via $e^+e^- \to a \gamma$ or $e^+e^- \to Z \gamma \to (a\gamma)\gamma$ and decaying into pairs of charged leptons. In the ideal case of zero background and perfect detection efficiency, we find that far detectors placed in the planned underground cavities or a large cuboid on the ground can enhance the sensitivity to long-lived pseudo-scalars at best moderately. On the other hand, the ILD itself is a perfect environment to search for long-lived particles, due to its excellent angular coverage and radial thickness. For long-lived particles produced with cross sections of a few picobarns, the ILD could probe lifetimes up to $300\,$ns, or proper decay lengths up to $100\,$m, in $250\,$fb$^{-1}$ of data. For axion-like particles produced through weak interactions, the ILC can reach an even higher sensitivity than searches for displaced vertices in meson decays at \belletwo. Our findings apply similarly to other proposed electron-positron experiments with a high angular coverage, such as the FCC-ee and CEPC.}

\begin{document}

\maketitle

\flushbottom

\section{Introduction}\label{SEC:introduction}
Current searches for long-lived particles (LLP) are often driven by opportunism. Existing particle colliders are found to be sensitive to LLPs with a certain range of masses and couplings, which is determined by the particle source and the detector geometry~\cite{Alimena:2019zri}. For the LHC, new search strategies~\cite{Alimena:2021mdu,Borsato:2021aum}, as well as several annex experiments like FASER~\cite{Feng:2017uoz,FASER:2019aik}, MATHUSLA~\cite{Curtin:2018mvb}, and CODEX-b~\cite{Gligorov:2017nwh} have been proposed to optimize and extend the reach for LLPs.\footnote{FASER is currently constructed and has demonstrated its capability to detect neutrinos with a test run~\cite{Ariga:2021vyy}.}

For the next high-energy particle collider, it is advisable to explore the discovery potential for LLPs and optimize the detector setup before construction. The key question to answer is: Where should an LLP detector ideally be placed and with what geometry?

In this work we answer this question for future high-energy electron-positron experiments. While we focus on the International Linear Collider (ILC)~\cite{Fujii:2017vwa}, most of our results can be transferred to the FCC-ee~\cite{FCC:2018evy} and CEPC~\cite{CEPCStudyGroup:2018ghi}. Previous analyses for future $e^+ e^-$ colliders have focused on LLPs produced via Higgs or $Z$ boson decays~\cite{Bauer:2018uxu,Alipour-Fard:2018lsf,Cheung:2019qdr} or sterile neutrinos~\cite{Antusch:2016vyf} and decaying in the main detector. The physics potential of far detectors has been explored at the FCC-ee or CEPC~\cite{Wang:2019xvx,Chrzaszcz:2020emg} and at future lepton colliders running at the $Z$ pole~\cite{Tian:2022rsi}. Alternatively, light LLPs produced from the electron or positron beams could be observed in beam dump experiments, as proposed for the ILC~\cite{Asai:2021ehn}.

We conduct the first study of realistic far detector options for the ILC with a center-of-mass energy of $\sqrt{s} = 250\,$GeV. A central aspect of our work is to compare the reach for LLPs at far detectors with the main detector ILD. We perform this comparison systematically for LLPs produced with different kinematic distributions. This allows us to determine the key features required for a detector to be sensitive to LLPs with specific production kinematics and decay lengths. For concreteness, we focus on axion-like particles (ALPs) $a$ with masses well below the weak scale, produced via $e^+ e^- \to a\gamma$ and $e^+ e^- \to Z\gamma \to (a\gamma)\gamma$ and decaying into pairs of displaced charged leptons. These two scenarios can be considered as benchmarks for a larger class of light new particles with similar production kinematics and decay lengths, which should lead to very similar results.

The second main aspect of our work is to compare our results for the ILC with low-energy electron-positron colliders, most notably the \belletwo experiment~\cite{Belle-II:2018jsg}. While both experiments benefit from a clean environment and an excellent angular coverage of the main detector, the dominant production channels and decay kinematics of an LLP are very different due to the different collision energies. If and how a high-energy electron-positron collider can extend the reach for LLPs beyond \belletwo is therefore a tricky question with a high impact on the discovery potential of future experiments.

This paper is structured as follows: In Sec.~\ref{SEC:alps}, we introduce the model used for our analysis and discuss the main production channels and kinematic distributions of axion-like particles at the ILC. In Sec.~\ref{SEC:detectors}, we design three far detectors at the ILC and analyze how their position and geometry affects the expected event rates of decaying LLPs. In Sec.~\ref{SEC:results}, we discuss the discovery potential for ALPs at the ILC far detectors in detail and compare it with the ILD. In Sec.~\ref{SEC:ilc-vs-belle2}, we compare the sensitivity at the ILC with ALPs from meson decays at \belletwo, before concluding in Sec.~\ref{SEC:conclusions}.

\section{Axion-like particles at the ILC}\label{SEC:alps}
Axion-like particles (ALPs) are pseudo-scalars that could originate as pseudo Nambu-Goldstone bosons from a chiral symmetry broken at a scale $\Lambda$ and whose couplings respect a shift symmetry $a \to a + c$. We focus on a benchmark model with ALPs coupling only to light leptons $\ell = \{e,\mu\}$ and weak gauge fields. At energies above the weak scale, $\mu_w$, the ALP couplings are described by an effective Lagrangian~\cite{Georgi:1986df}
\begin{align}\label{eq:lagrangian}
    \mathcal{L}_{\rm eff}(\mu > \mu_w) &=  \frac{c_{\ell\ell}}{2}\,\frac{\partial^{\mu} a}{f_a} \big(\bar{\ell} \gamma_{\mu} \gamma_5 \ell\big) + c_{WW}\,\frac{\alpha_2}{4\pi}\,\frac{a}{f_a}\,W_{\mu\nu}^\tau \widetilde{W}^{\mu\nu}_\tau\,,
\end{align}
where $c_{\ell\ell}$ denotes the ALP coupling to leptons, and $c_{WW}$ is the coupling to weak gauge bosons with field strength tensor $W_{\mu\nu}^\tau$. The latter is normalized to the weak gauge coupling $\alpha_2 = \alpha/s^2_w$, with the fine structure constant $\alpha$ and the sine of the weak mixing angle, $s_w = \sin\theta_w$. The effective theory is valid at energies up to a cutoff scale $\Lambda = 4\pi f_a$, where it is to be completed by a full model. For the QCD axion, the scale $f_a$ is related to the axion decay constant.

\begin{figure}[t!]
  \centering
    \raisebox{0.4cm}{\includegraphics[width=0.35\textwidth]{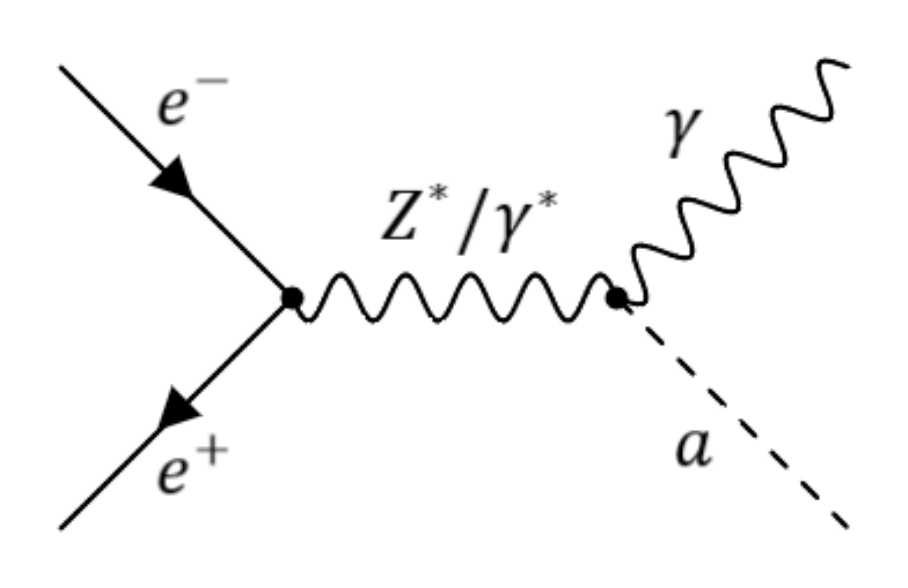}}\hspace*{1.6cm}
    \includegraphics[width=0.28\textwidth]{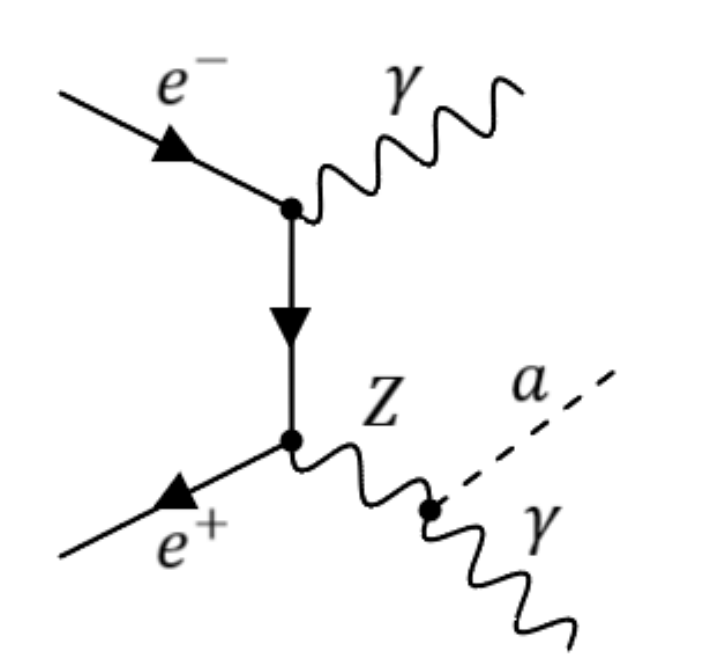}\\
    \caption{Feynman diagrams for ALP production via $e^+e^- \to a \gamma$ (left) and $e^+e^- \to Z\gamma \to (a \gamma)\gamma$ (right).}
    \label{fig:feynman}
\end{figure}

Below the scale of electroweak symmetry breaking, the ALP interactions with photons and $Z$ bosons read
\begin{align}\label{eq:lagrangian-gauge}
    \mathcal{L}_{\rm eff}(\mu < \mu_w) & \supset \frac{\alpha}{4\pi}\frac{a}{f_a}\left(c_{\gamma \gamma}\,F_{\mu\nu}\widetilde{F}^{\mu\nu} + 2\frac{c_{\gamma Z}}{s_w c_w}F_{\mu\nu}\widetilde{Z}^{\mu\nu} + \frac{c_{ZZ}}{s^2_w c^2_w}Z_{\mu\nu}\widetilde{Z}^{\mu\nu}\right),
\end{align}
with the field strength tensors of the photon, $F_{\mu\nu}$, and the $Z$ boson, $Z_{\mu\nu}$, and the couplings
\begin{align}
c_{\gamma\gamma} = c_{WW},\qquad c_{\gamma Z} = c_w^2\,c_{WW},\qquad c_{ZZ} = c_w^4\,c_{WW}\,.
\end{align}
At the ILC running at a center-of-mass energy of $\sqrt{s} = 250\,$GeV, ALPs \sw{with masses below the weak scale} can be efficiently produced through the coupling $c_{WW}$\footnote{ALP production from electrons is strongly suppressed by the small electron mass.} via two main production channels
\begin{align}
e^+e^- \to a \gamma\qquad \text{and}\qquad e^+e^- \to Z\gamma \to (a \gamma)\gamma\,.
\end{align}
Illustrative Feynman diagrams are shown in Fig.~\ref{fig:feynman}.
 In $e^+e^- \to a \gamma$~\cite{Mimasu:2014nea,Jaeckel:2015jla}, the ALP is directly produced in association with a photon; the cross section scales as \sw{$\sigma(e^+e^- \to a \gamma) \sim c_{\gamma Z}^2$ or $c_{\gamma\gamma}^2$.} In $e^+e^- \to Z\gamma \to (a \gamma)\gamma$, the production proceeds through an on-shell $Z$ boson, which decays with a branching ratio $\Br(Z \to a\gamma)\sim c_{\gamma Z}^2$, so that
\begin{align}
\sigma(e^+e^- \to (a \gamma)\gamma) \equiv \sigma(e^+e^- \to Z \gamma)\,\Br(Z\to a \gamma)\,.
\end{align}
At $\sqrt{s} = 250\,$GeV and with unpolarized beams, the ALP production rates are given by
\begin{align}\label{eq:prod-xs}
\sigma(e^+e^- \to a \gamma) = 298 \left(\frac{c_{WW}}{f_a\,[\text{TeV}]}\right)^2\text{fb},\quad\ \sigma(e^+e^- \to (a \gamma)\gamma) = 144\left(\frac{c_{WW}}{f_a\,[\text{TeV}]}\right)^2\text{fb}\,.
\end{align}
ALP production through weak boson fusion, $e^+ e^- \to a e^+ e^-$ or $e^+ e^- \to a \nu \bar{\nu}$, is smaller by about an order of magnitude at $\sqrt{s} = 250\,$GeV, \sw{but could become relevant at higher collision energies. The production of sub-GeV ALPs through meson decays, notably $B\to K a$ and $K\to \pi a$, is suppressed by several orders of magnitude due to the small branching ratio, see e.\,g.~\cite{Ferber:2022rsf}.}

\sw{Sub-GeV ALPs could also be produced in the material around the ILD via the Primakoff process, thus enhancing the ALP rate in the far detectors compared to the ILD.\footnote{This effect has been studied at the LHC for the far forward facility FASER~\cite{Feng:2018pew}.} At long lifetimes, however, we do not expect a significant enhancement, due to the shorter distance between production and detection point and due to the lower boost of Primakoff-produced ALPs. A dedicated investigation of this effect is nevertheless interesting; we leave it for future work.}

\sw{Very light ALPs could finally be produced from photon conversion in the magnetic field inside the beam pipe. We do not consider this option and focus on ALPs with masses between the MeV scale and the weak scale.}

In our numerical analysis we set
\begin{align}
\frac{c_{WW}}{f_a} = \frac{\sw{0.3}}{\rm TeV} \qquad \text{and} \qquad m_a = 300\,\text{MeV},
\end{align}
unless specified otherwise, \sw{and keep the lepton coupling $c_{\ell\ell}$ as a free parameter. Varying the coupling changes the production rate as $(c_{WW}/f_a)^2$, see~\eqref{eq:prod-xs}. Varying the mass has almost no impact on the production rate, as long as $m_a < m_Z$.}

The lifetime and decay channels of the ALP, however, depend strongly on its mass and couplings. In our \sw{benchmark scenario, the ALP decays mostly into muon pairs; for $c_{\ell\ell}/c_{WW} \lesssim 0.002$ the decay to photons starts dominating.}\footnote{We also allow for decays to electrons, which however are strongly suppressed and negligible for ALPs with masses above the di-muon threshold.} The ALP decay width to muons is
\begin{align}\label{eq:width}
\Gamma(a\to \mu^+\mu^-) = \frac{m_a m_\mu^2}{8\pi}\left(\frac{c_{\ell\ell}}{f_a}\right)^2\sqrt{1-\frac{4 m_\mu^2}{m_a^2}}\,,
\end{align}
where $m_\mu$ is the muon mass. For $m_a = 300\,$MeV \sw{and $c_{\ell\ell}/c_{WW} \gtrsim 0.002$}, the proper decay length of the ALP is given by
\begin{align}\label{eq:decay-length}
c\tau_a = \frac{c}{\Gamma(a\to \mu^+\mu^-)} \approx 50\,\left(\frac{f_a\,[\text{TeV}]}{c_{\ell\ell}}\right)^{2} \mu\text{m}.
\end{align}
The decay length observed in the laboratory frame, $d = (\beta\gamma)_a c\tau_a$, is related to the proper decay length $c\tau_a$ by the boost $(\beta\gamma)_a$. Most of our analysis relies on the production kinematics and decay length of the ALP. Using \eqref{eq:prod-xs} and \eqref{eq:decay-length}, the results can be translated to other light (pseudo-)scalars with different decay modes, as long as the production kinematics are similar.

\begin{figure}[t!]
    \centering
    \includegraphics[width=0.5\textwidth]{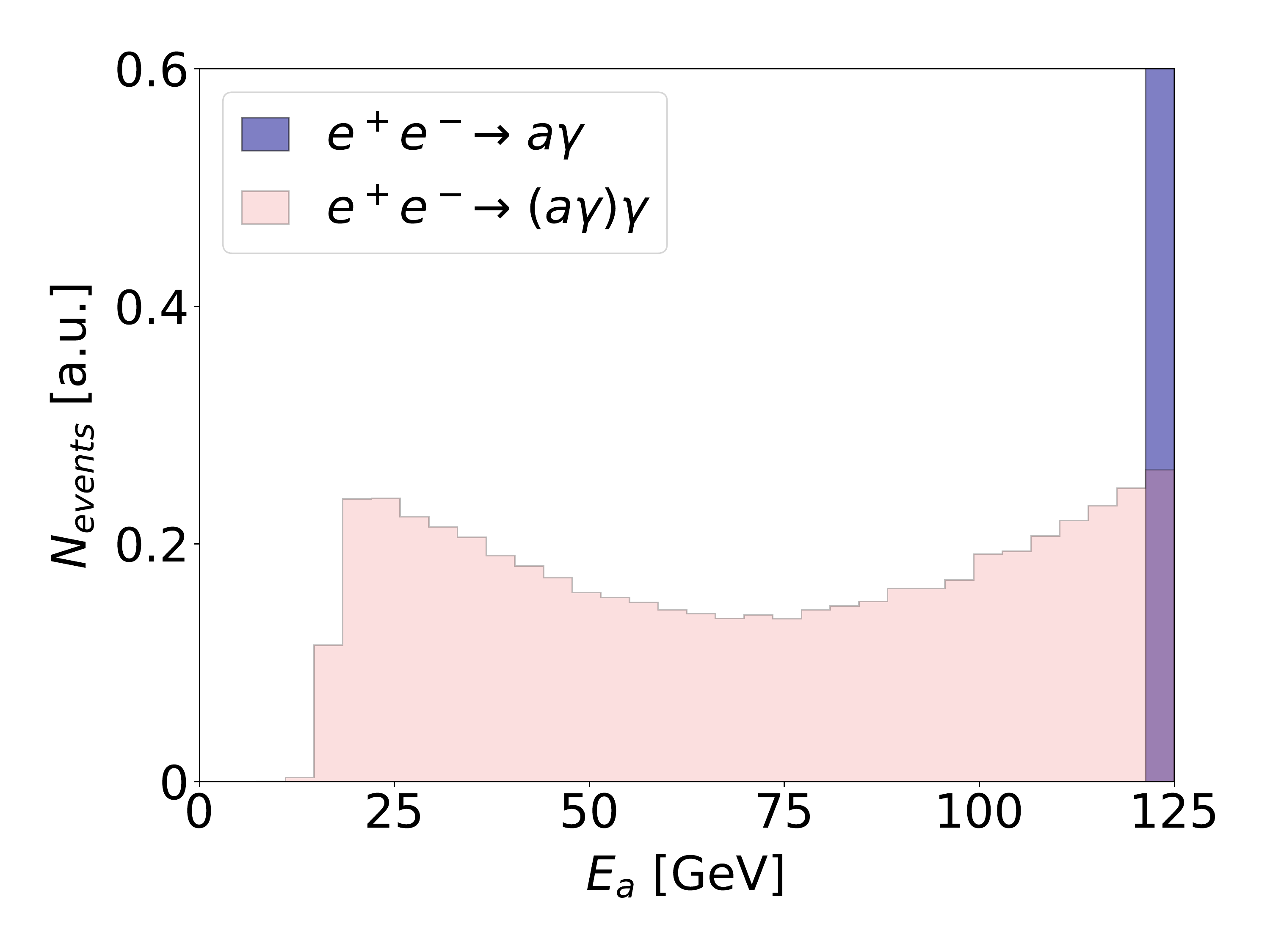}\hfill
    \includegraphics[width=0.5\textwidth]{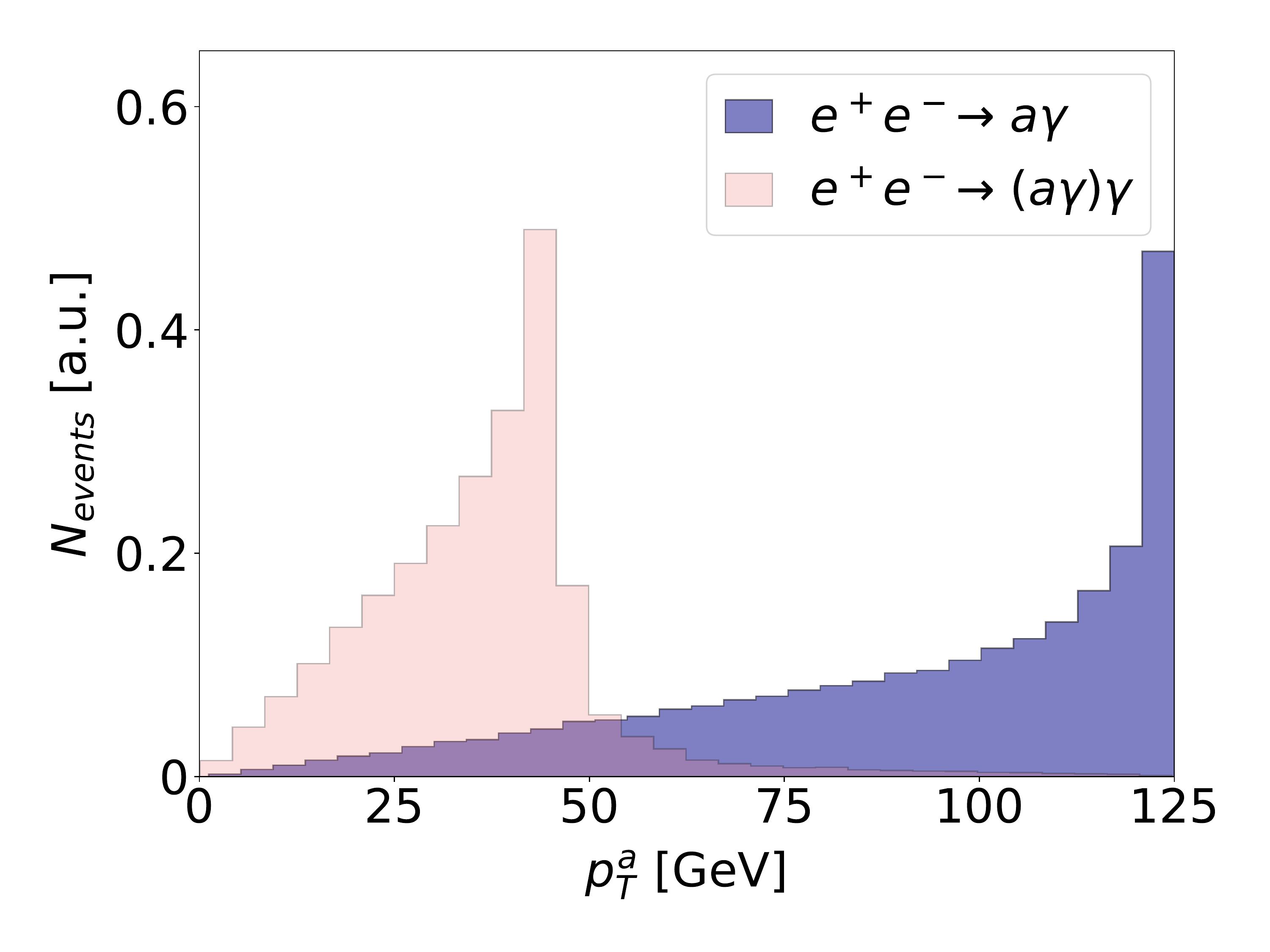}\\
     \includegraphics[width=0.5\textwidth]{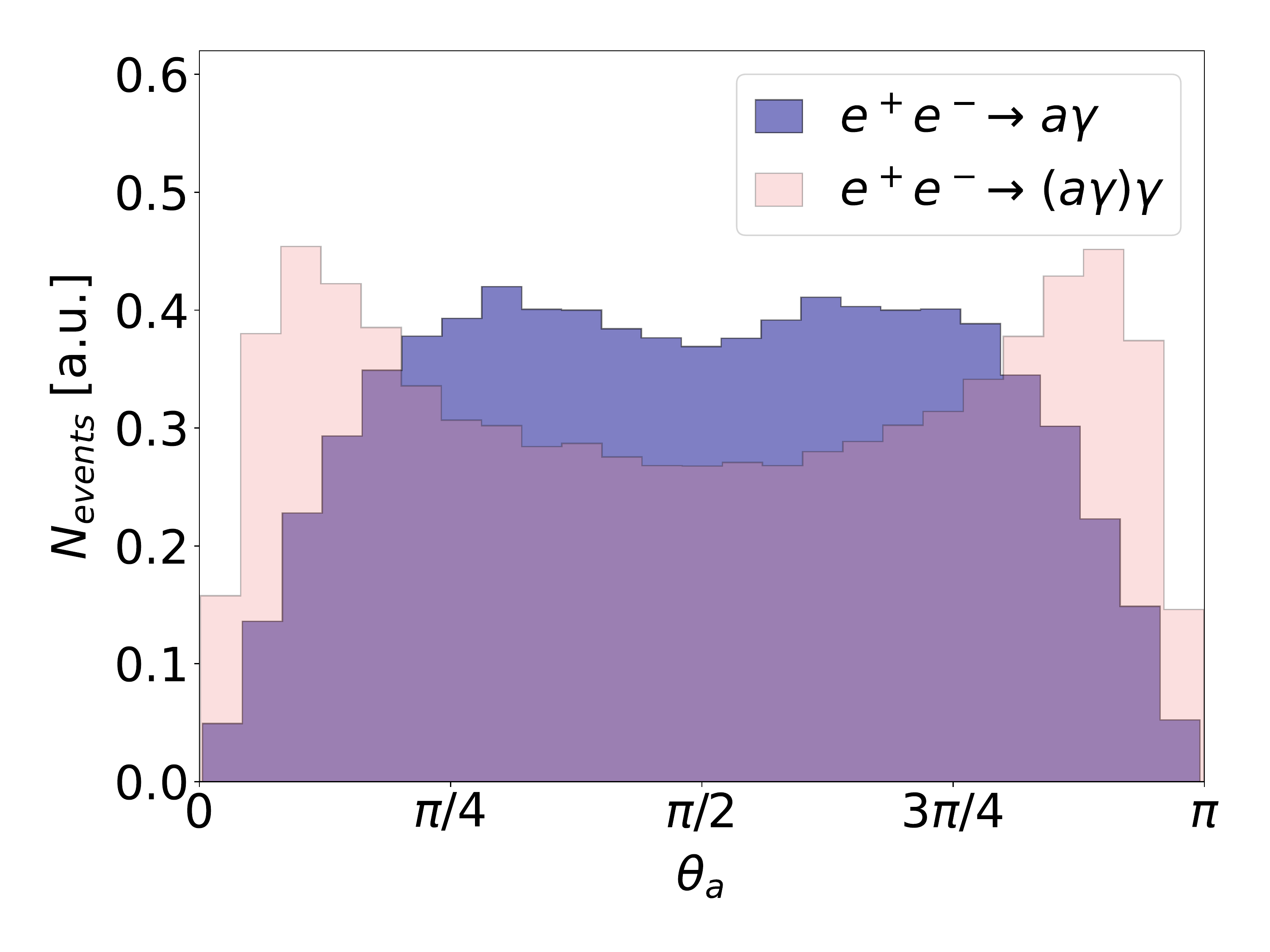}
    \caption{Kinematic distributions of an ALP with $m_a = 300\,$MeV, produced via $e^+e^- \to a \gamma$ (dark blue) and $e^+e^- \to (a \gamma) \gamma$ (light red) at the ILC with $\sqrt{s} = 250\,$GeV: energy $E_a$ (upper left), transverse momentum $p_T^a$ (upper right), and scattering angle $\theta_a$ (lower).}
    \label{fig:kinematics}
\end{figure}

The rate of ALP decays inside a detector depends not only on the ALP decay length, but also on its kinematic distributions. In Fig.~\ref{fig:kinematics}, we show the differential distributions of the ALP in energy $E_a$, transverse momentum $p_T^a$, and scattering angle $\theta_a$ off the electron beam for the two production processes $e^+e^- \to a \gamma$ and $e^+e^- \to (a \gamma) \gamma$.\footnote{\sw{These distributions have been obtained from simulations using {\tt MadGraph5\_aMC@NLO}~\cite{Alwall:2014hca,Frederix:2018nkq}. See Sec.~\ref{SEC:results} for details.}} All kinematic variables are defined in the laboratory frame, which corresponds with the $e^+ e^-$ center-of-mass frame.

For $e^+e^- \to a \gamma$, the energy of the two final-state particles is peaked at $E_a = \sqrt{s}/2 = 125\,$GeV. This is also reflected in the transverse momentum spectrum of the ALP, which peaks near $p_T^a \approx 125\,$GeV. \sw{The angular distribution
\begin{align}
\frac{d\sigma(e^+e^- \to a \gamma)}{d\theta_a} \propto \sin\theta_a(1 + \cos^2\theta_a)
\end{align}
causes the majority of ALPs to be emitted in the central region perpendicular to the beam axis.}

For  $e^+e^- \to (a \gamma) \gamma$, the kinematic distributions look very different, due to the intermediate on-shell $Z$ boson. The cross section for $e^+ e^- \to Z \gamma$ features a collinear enhancement in the forward region, see e.\,g.~(7) in~\cite{An:2015pva}, so that the $Z$ boson is emitted along the beam line. Due to momentum conservation, the ALP in $Z\to a\gamma$ is produced either in or against the direction of the $Z$. In the former case the ALP carries the energy of the $Z$ boson, $E_a \approx 125\,$GeV, while in the latter case the ALP carries just enough momentum to balance the recoil of the two photons. This explains the enhancement in the energy spectrum near the kinematic endpoints, as well as the enhancement in the forward and backward direction in the angular distribution.

In summary, the ALP distributions from the two production processes have different features:
\begin{align}
e^+e^- \to a \gamma: & \ \quad (\beta\gamma)_a \approx E/m_a \approx 400\quad \text{(central)}\\\nonumber
e^+e^- \to (a \gamma) \gamma: & \ \quad 60 \lesssim (\beta\gamma)_a \lesssim 400\qquad\ \text{(forward)}.
\end{align}

From these considerations it becomes clear that the number of ALP decays within a detector with limited angular coverage and thickness depends significantly on the production kinematics and the placement of the detector with respect to the production point.

\sw{In other models, long-lived particles might be produced through alternative processes, for instance, dark scalars from Higgs decays~\cite{Alipour-Fard:2018lsf,Cheung:2019qdr,Wang:2019xvx} or sterile neutrinos from $t$-channel $W$ boson exchange~\cite{Cheung:2019qdr,Wang:2019xvx}. As long as their angular distribution and boost are comparable to those for ALPs, we expect a similar  reach in lifetime at far detectors. Our expectation is based on the fact that the decay rate within a certain detector volume relies on the kinematics and the lifetime of the particle, not on the exact details of its couplings or production channels. Such similarities have previously been observed in a study of far detectors at the $e^+ e^-$ experiment Belle II, similar in spirit, which compares models with ALPs, dark scalars and sterile neutrinos~\cite{Dreyer:2021aqd}. The reach of far detectors in that study are largely independent of the model details, as long as the direction and boost of the long-lived particles are comparable.}

\section{Far detector options}
\label{SEC:detectors}
The detection of a long-lived particle crucially depends on its decay probability at a distance from the production point. In this section, we review how the decay probability scales with the geometry and the position of a detector. Based on these considerations, we design three realistic far detectors at the ILC and compare their expected acceptance with the proposed main detector ILD.

\paragraph{Decay probability} The probability for a particle $a$, produced in the direction $\vec{r}_a$ with decay length $d_a$, to decay within a distance $r \in [r_a^{\rm in},r_a^{\rm out}]$ from its production point is given by
\begin{align}\label{eq:decay-prob}
\mathds{P}_a(d_a;\vec{r}_a) = \exp\left(-\frac{r_a^{\rm in}}{d_a}\right) - \exp\left(-\frac{r_a^{\rm out}}{d_a}\right).
\end{align}
In practice, $r_a^{\rm in}$ and $r_a^{\rm out}$ are the positions at which the particle's trajectory intersects with the detector boundaries. For a sample of $N$ particles, the average decay probability is obtained as
\begin{align}
\langle \mathds{P} \rangle = \frac{1}{N} \sum_{a=1}^N\, \mathds{P}_a\left(d_a;\vec{r}_a\right).
\end{align}
For particles with very long decay lengths, the average decay probability can be approximated by
\begin{align}\label{eq:decay-prob-approx}
\langle d\rangle \gg  r^{\rm in}, \langle r \rangle: \quad \langle \mathds{P} \rangle \approx \frac{\Omega}{4\pi}\frac{\langle r \rangle}{\langle d \rangle}\,,
\end{align}
where $\Omega$ is the solid angle covered by the detector, $\langle r \rangle$ is the average radial thickness of the detector, and $\langle d \rangle$ is the particle's decay length averaged over its boost. As long as the kinematic distribution is sufficiently isotropic, the probability of very long-lived and/or highly boosted particles to decay inside the detector scales linearly with the detector dimensions $\Omega \cdot \langle r \rangle$. Below we will use $\Omega \cdot \langle r \rangle$ as a measure of acceptance when we consider different detector geometries.

At the ILC, we calculate the number of particles $a$ that decay within a certain detector volume as~\footnote{\sw{The approximation is valid for isotropic particle production and decay. In our numerical analysis we perform a full Monte Carlo simulation without making approximations.}}
\begin{align}\label{eq:na}
N_a = \mathcal{L}\, \frac{1}{N}\sum_{a=1}^N \sigma\left(e^+ e^- \to a\left(d_a;\vec{r}_a\right) X\right) \mathds{P}_a\left(d_a;\vec{r}_a\right) \approx \mathcal{L}\, \sigma\left(e^+ e^- \to a X\right) \langle \mathds{P}\rangle\,,
\end{align}
\sw{that is, we weigh the probability that particle $a$ decays within the detector with the differential production cross section, sum over all events $a = 1,\dots N$ and normalize to the total number of $a$ decays, $N$. The integrated luminosity is set to $\mathcal{L} = 250\,$fb$^{-1}$ in our analysis.}

\paragraph{Near and far detectors} Two concepts for near detectors at the ILC have been developed, named ILD and SiD~\cite{ILC:2007vrf,ILD:2019kmq}. We focus on the ILD, which is designed to be bigger and a priori more sensitive to particle decays away from the $e^+ e^-$ collision point. For our numerical analysis, we consider two parts of the detector that are well suited to reconstruct displaced vertices of charged particles: the multi-layer pixel vertex detector (VTX), which promises a high-precision vertex reconstruction, and the time projection chamber (TPC), which allows for tracking and timing over a larger volume. Taken together, the VTX and TPC parts of the ILD form a cylindrical decay volume for LLPs centered around the $e^+ e^-$ collision point, with
\begin{itemize}
\item ILD: $\quad z \in [-235,235]\,$cm, $\quad \rho \in [0.6,180.8]\,$cm,  $\quad \theta \in [8,172]\degree .$
\end{itemize}
Here $z$ is the $e^-$ beam direction, $\rho$ is the radial distance from the beam axis, and $\theta$ is the polar angle around the beam. The ILD has an excellent angular coverage, so that almost all LLPs decaying within the cylindrical shell with $\rho \in [0.6,180.8]\,$cm can be detected.

\begin{figure}[t!]
    \centering
    \includegraphics[width=0.49\textwidth]{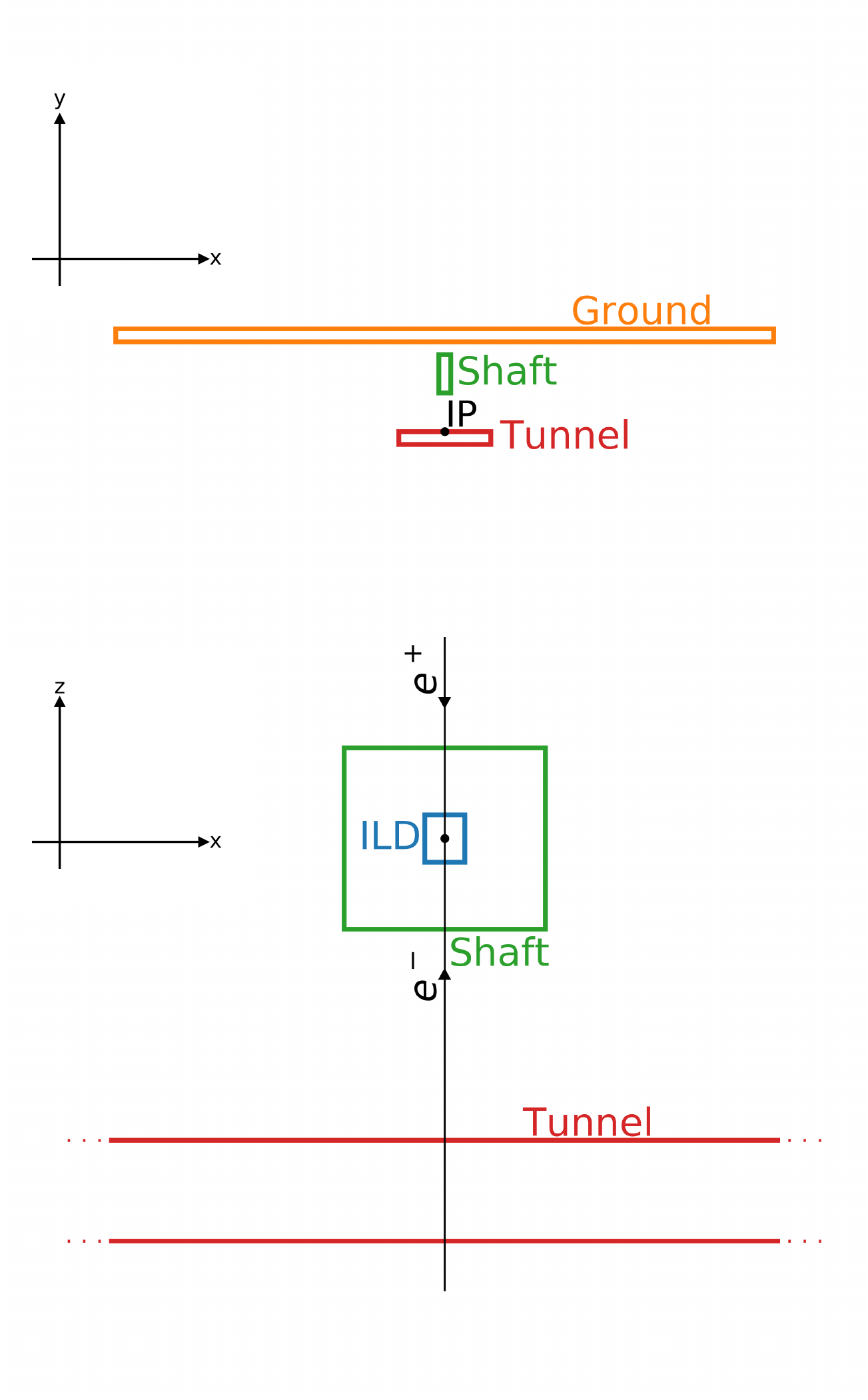}
    \caption{Far detector options around the ILC interaction point (IP). Shown are a side view (left) and top view (right) of the projected far detectors in the Shaft (S, blue), in the Tunnel (T, purple), and on the Ground (G, red), as well as the main detector ILD (green). The Ground detector is centered around $(x,z) = (0,0)$ and is too large to appear in the top view.}
    \label{fig:detectors}
\end{figure}

In addition to the ILD, we consider three options for far detectors that could be installed in planned underground cavities around the ILC detector hall or on the ground. One possible site is the vertical shaft above the collision point, which will be used to lower the main ILD and SiD detectors into the detector hall. A second possibility is to place a far detector inside the  access tunnel that surrounds the detector hall. As a third option, we consider a large detector placed on the ground above the detector hall. We design three cuboid detectors with the following extensions:
\begin{itemize}
\item Shaft (S): $\,\qquad 18\times 30\times 18\,$m, centered around $(0,45,0)\,$m
\item Tunnel (T): $\ \quad 140\times 10\times 10\,$m, centered around $(0,-5,-35)\,$m
\item Ground (G): $\quad 1000\times 10\times 1000\,$m, centered around $(0,75,0)\,$m\,.
\end{itemize}
The positioning $(x,y,z)$ of the detectors in the reference frame around the $e^+e^-$ interaction point is illustrated in Fig.~\ref{fig:detectors}.
 As suggested by \eqref{eq:decay-prob-approx}, the detector acceptance for long-lived particles is determined by the product of angular coverage and average radial thickness, $\Omega \cdot \langle r \rangle$.\footnote{We calculate $\langle r \rangle$ as the arithmetic mean of the distances $r_a^{\rm out} - r_a^{\rm in}$ in a large sample of $N$ particle trajectories that intersect the detector volume, $\langle r \rangle = N^{-1}\sum_{a=1}^N (r_a^{\rm out} - r_a^{\rm in})$.} In Tab.~\ref{tab:geometries}, we summarize these properties for the four proposed detector geometries.
\begin{table}[t]
\renewcommand{\arraystretch}{1.3}
\begin{center}
\begin{tabular}{r||c|c|c|c}
\toprule
 & ILD & Shaft & Tunnel & Ground \\
\midrule
$\Omega/(4\pi)$\phantom{i} & 0.999 & 0.026 & 0.046 & 0.44 \\
$\langle r \rangle$ [m]\phantom{i} & 2.2 & 16 & 11 & 23 \\
\hline
$\Omega \cdot \langle r \rangle$ [sr\,m]\phantom{i} & 27 & 5 & 6 & 126 \\
\bottomrule
\end{tabular}
\end{center}
  \caption{Geometric properties of the ILD and the three projected far detectors Shaft, Tunnel and Ground: angular coverage $\Omega/(4\pi)$; average radial thickness $\langle r \rangle$; and measure of acceptance $\Omega \cdot \langle r \rangle$. \label{tab:geometries}}
\end{table}
 Based on the measure of acceptance, $\Omega \cdot \langle r \rangle$, we expect that the ILD will detect larger numbers of long-lived particles compared to the Shaft and Tunnel detectors, which suffer from a small angular coverage. With the Ground detector option, the acceptance increases by roughly a factor of 4 compared to the ILD. This is due to the huge extent of the Ground detector, which compensates for the loss in angular coverage at a large distance from the collision point. \sw{Such a huge detector seems technically unrealistic. We choose this extreme design to demonstrate the maximum gain that could be achieved with any realizable large surface detector.}
 
Notice that these considerations apply for isotropically produced particles. For particles emitted mostly perpendicular to the beam line, the fraction of detected events at all three far detectors increases. At the ILD, in turn, the detection rate depends little on the angular distribution, due to the almost perfect angular coverage.

\section{ILC discovery potential}
\label{SEC:results}
Based on the results of Sections~\ref{SEC:alps} and \ref{SEC:detectors}, we determine the reach of the three proposed far detectors for long-lived ALPs and compare it with the ILD. To this end, we have simulated $100\,$k events for each of the processes $e^+ e^- \to a\gamma$ and $e^+ e^- \to Z\gamma \to (a\gamma)\gamma$ with unpolarized initial leptons, using {\tt MadGraph5\_aMC@NLO}~\cite{Alwall:2014hca,Frederix:2018nkq}. With these samples, we then analyze the sensitivity of all four detectors to long-lived ALPs using our own analysis framework written in {\tt Python} \cite{10.5555/1593511} and {\tt Jupyter}~\cite{soton403913}.

\begin{figure}[t!]
    \centering
    \includegraphics[width=0.50\textwidth]{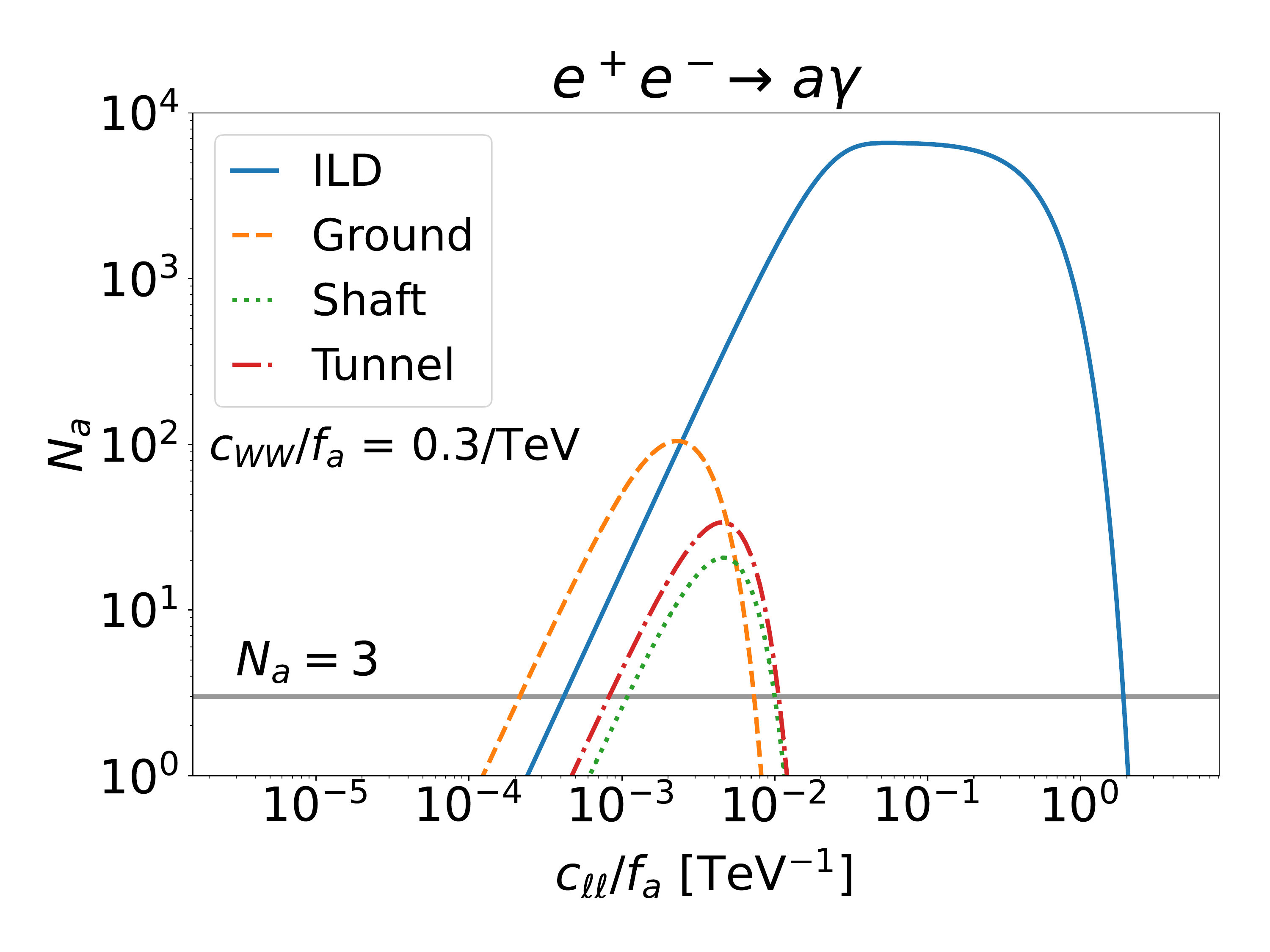}\hspace*{0.1cm}
    \includegraphics[width=0.50\textwidth]{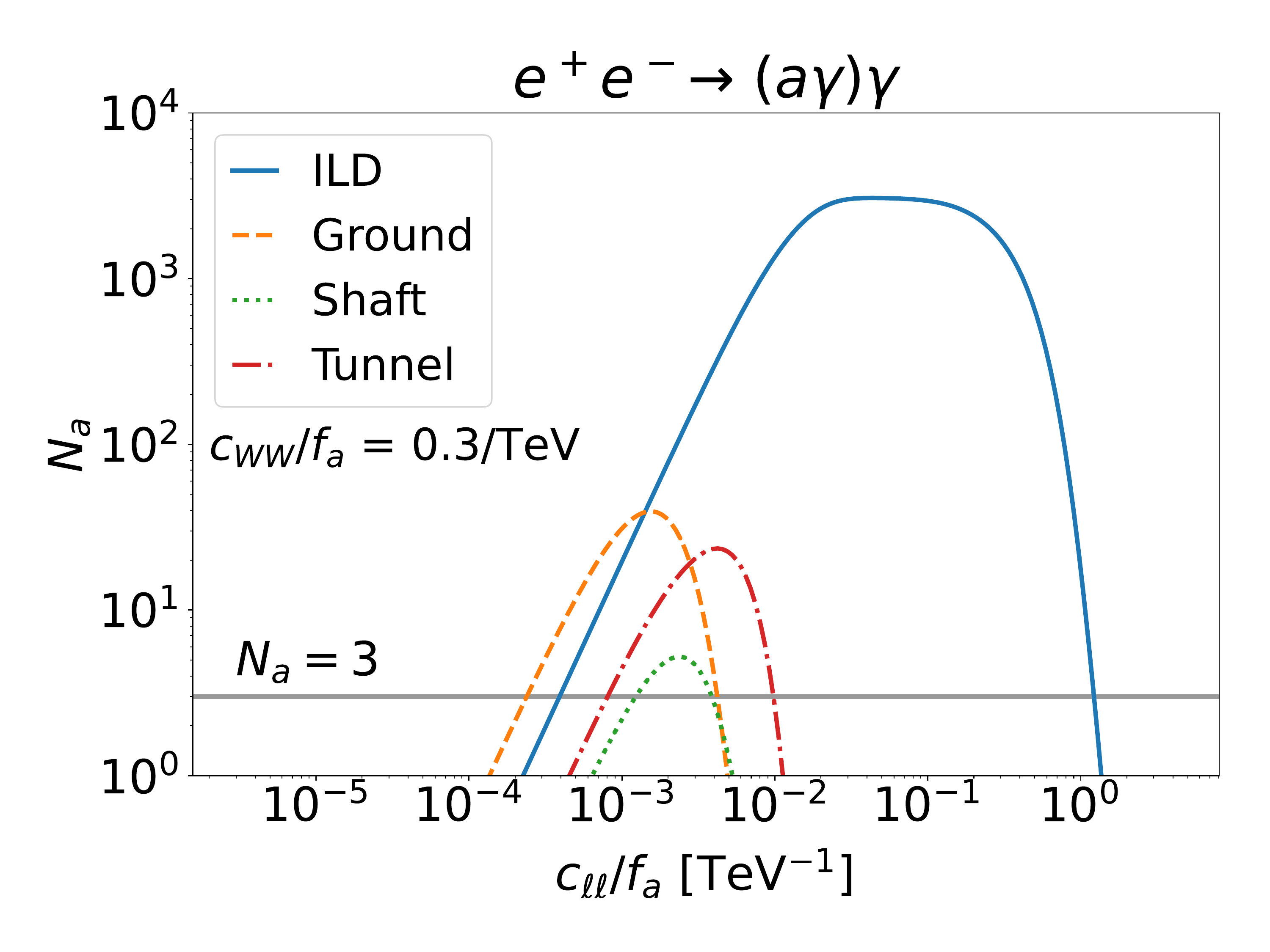}
    \caption{Number of ALPs, $N_a$, decaying within various ILC detectors, as a function of the effective coupling to leptons, $c_{\ell\ell}/f_a$, fixing $c_{WW}/f_a = \sw{0.3}/$TeV and $m_a = 300\,$MeV. The event rates correspond to the production channels $e^+e^- \to a\gamma$ (left) and $e^+e^- \to Z\gamma \to (a\gamma)\gamma$ (right) at $\sqrt{s} = 250\,$GeV and with $\mathcal{L} = 250\,$fb$^{-1}$. Shown are predictions for the ILD (blue, \sw{plain}) and far detectors placed in the Shaft (green, \sw{dotted}), in the Tunnel (red, \sw{dot-dashed}) and on the Ground (orange, \sw{dashed}).}
    \label{fig:ndec-cll}
\end{figure}

\paragraph{Theoretical sensitivity to long-lived ALPs} In Fig.~\ref{fig:ndec-cll}, we show the expected number of ALPs, $N_a$, as defined in~\eqref{eq:na}, for the ILD (blue), as well as the far detectors Shaft (green), Tunnel (red), and Ground (orange). We display the event rates as a function of the coupling $c_{\ell\ell}/f_a$ which determines the ALP's decay width~\eqref{eq:width}, for fixed $c_{WW}/f_a = \sw{0.3}/$TeV.
Going from larger to smaller couplings $c_{\ell\ell}/f_a$, the event rates in all detectors increase exponentially as $\langle \mathds{P} \rangle \sim \exp(-r^{\rm in} (c_{\ell\ell}/f_a)^2)$ and eventually drop as $\langle \mathds{P} \rangle \sim \langle r \rangle (c_{\ell\ell}/f_a)^2$, see~\eqref{eq:decay-prob} and~\eqref{eq:decay-prob-approx}. For the ILD, the event rate reaches its maximum at larger values of $c_{\ell\ell}/f_a$, \emph{i.\,e.}, at smaller decay lengths $\langle d \rangle \sim (c_{\ell\ell}/f_a)^{-2}$ than for the far detectors, because the ILD is located closer to the production point. Moreover, the ILD curve features a plateau around $r^{\rm in} < \langle d \rangle < \langle r \rangle$, which is not observed for the far detectors. Indeed, the ILD is thick compared to its distance from the production point, $\langle r \rangle \gg r^{\rm in}$, so that the exponential increase of the event rate with $\langle d \rangle$ saturates before it decreases at large decay lengths. For a thinner detector at the same location or for the same detector located further away from the production point, the plateau becomes less broad and eventually disappears. This is the case for the far detectors, where $\langle r \rangle \lesssim r^{\rm in}$ and the event rates peak around $\langle d\rangle \gg r^{\rm in}$, see also Ref.~\cite{Tian:2022rsi}.

Besides the detector position and geometry, the absolute event rate also depends on the production cross sections from~\eqref{eq:prod-xs}. For a specific production process, the predicted rate for the ILD is much larger than at the far detectors, except for ALPs with small couplings or long lifetimes. Moreover, for the ILD the shape of the distribution is largely process-independent. Both features are due to the excellent angular coverage of the ILD around the production point. In turn, at the far detectors the sensitivity to ALPs with different lifetimes depends on the production kinematics. In particular, ALPs from $e^+ e^- \to (a\gamma)\gamma$ are boosted along the beam direction, see Fig.~\ref{fig:kinematics}. Since the event rates depend on the ratio of boost and coupling, $d \sim (\beta\gamma)_a(c_{\ell\ell}/f_a)^{-2}$, a larger boost is compensated by a larger coupling and vice versa. This effect leads to enhanced event rates at large couplings for the Tunnel detector (placed in the beam direction) and reduced rates at the Shaft detector (located centrally and perpendicular to the beam).

To illustrate the expected sensitivity to an ALP signal, in Fig.~\ref{fig:ndec-cll} we indicate event rates of $N_a = 3$ as horizontal dashed lines. In the ideal case of zero background, the observation of 3 events would correspond to excluding the no-signal hypothesis at 95\% CL, assuming Poisson statistics and 100\% reconstruction efficiency. Of course, such a scenario is unrealistic, \sw{given that the detectability of the ALP decay products can strongly depend on whether they can be resolved or how much distance they penetrate in the detector. Optimizing the detection technique, for instance by triggering on prompt associated particles in the ILD or by using timing information, would be worth a study on its own. Here we confine ourselves to comparing} the sensitivity of the respective detector \sw{geometries}, assuming that the background rate and reconstruction efficiency is similar for all detectors. In Tab.~\ref{tab:cll-bounds}, we show the smallest ALP couplings $c_{\ell\ell}/f_a$ that can be probed at the ILD and at the far detectors. Smaller couplings imply a higher sensitivity to long-lived ALPs.
\begin{table}[t!]
\renewcommand{\arraystretch}{1.3}
    \begin{center}
        \begin{tabular}{c||c|c|c|c}
            \toprule
            $c_{\ell\ell}/f_a\ [10^{-4}/\text{TeV}]$ & ILD & Shaft & Tunnel & Ground \\
            \midrule
            $e^+e^-\to a\gamma$ & $4.2$ & $10.8$ & $8.2$ & $2.1$ \\
            $e^+e^-\to (a\gamma)\gamma$ & $3.9$ & $12.3$ & $8.0$ & $2.4$ \\
            \bottomrule
        \end{tabular}
    \end{center}
  \caption{Expected sensitivity to the ALP coupling to leptons, $c_{\ell\ell}/f_a$, for two production processes with fixed $c_{WW}/f_a = \sw{0.3}/\text{TeV}$, at the ILD and three far detectors. The values correspond to observing $N_a = 3$ signal events from an ALP with mass $m_a = 300\,$MeV in an ideal experiment with zero background, which would exclude the no-signal hypothesis at the 95\% CL. Small couplings $c_{\ell\ell}/f_a$ indicate a high sensitivity to ALPs with long lifetimes.\label{tab:cll-bounds}}
\end{table}
 \sw{A large} Ground detector can improve the sensitivity by \sw{at most} a factor 2 compared to the ILD. \sw{Notice that} the improvement is much smaller than naively expected from the acceptance measures in Tab.~\ref{tab:geometries}. The reason is that the approximation $\langle \mathds{P} \rangle \sim \Omega\cdot \langle r \rangle$ from~\eqref{eq:decay-prob-approx} does not apply for the ILD, because the condition $\langle d\rangle \gg r^{\rm in}, \langle r \rangle$ is not fulfilled for the processes we consider. The Shaft and Tunnel detectors collect smaller event rates than the ILD. They might only be useful additions in the case of better background rejection or higher reconstruction efficiency than at the ILD.

\begin{figure}[t!]
    \centering
    \includegraphics[width=0.5\textwidth]{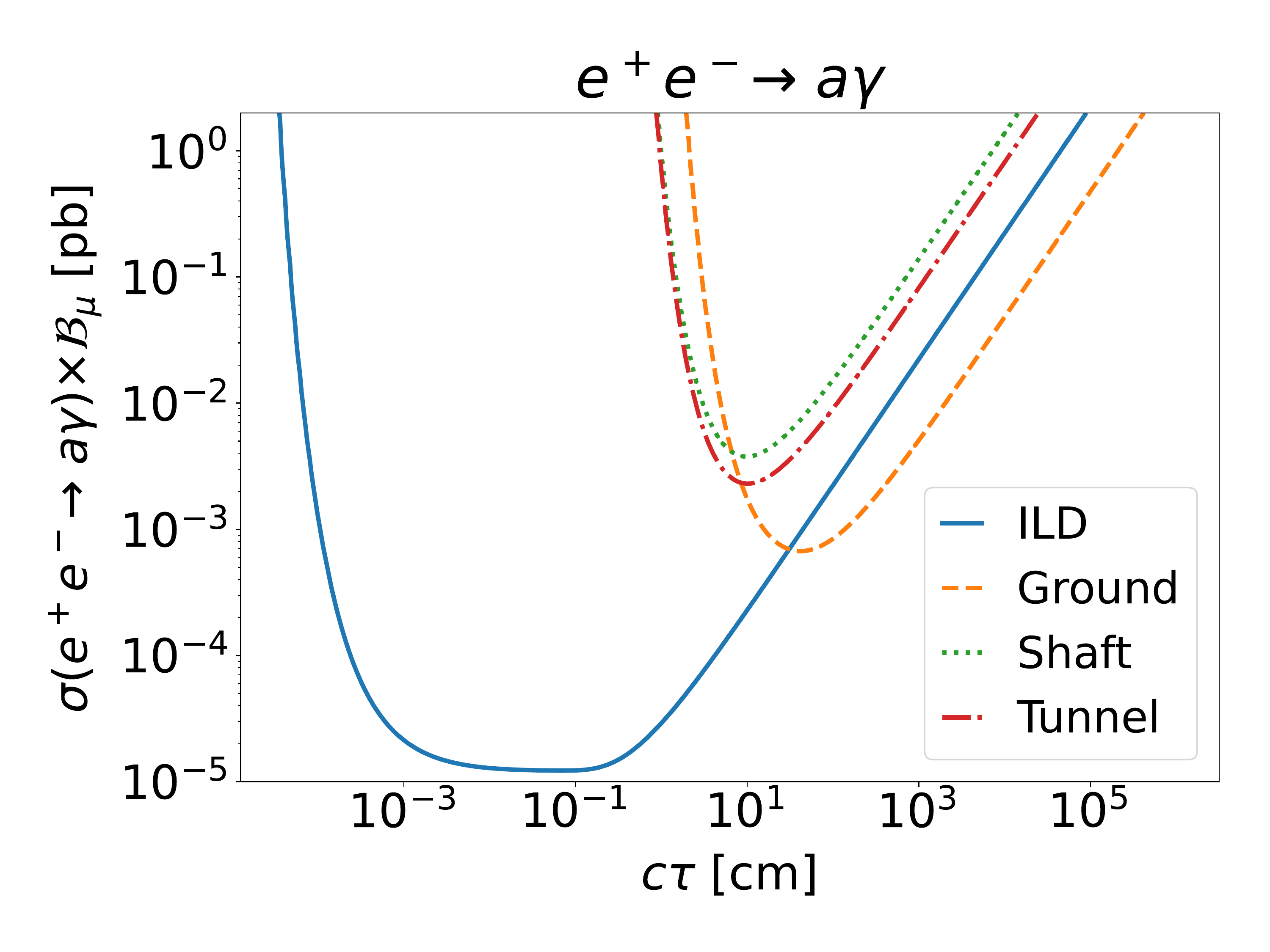}\hspace*{0.1cm}
    \includegraphics[width=0.5\textwidth]{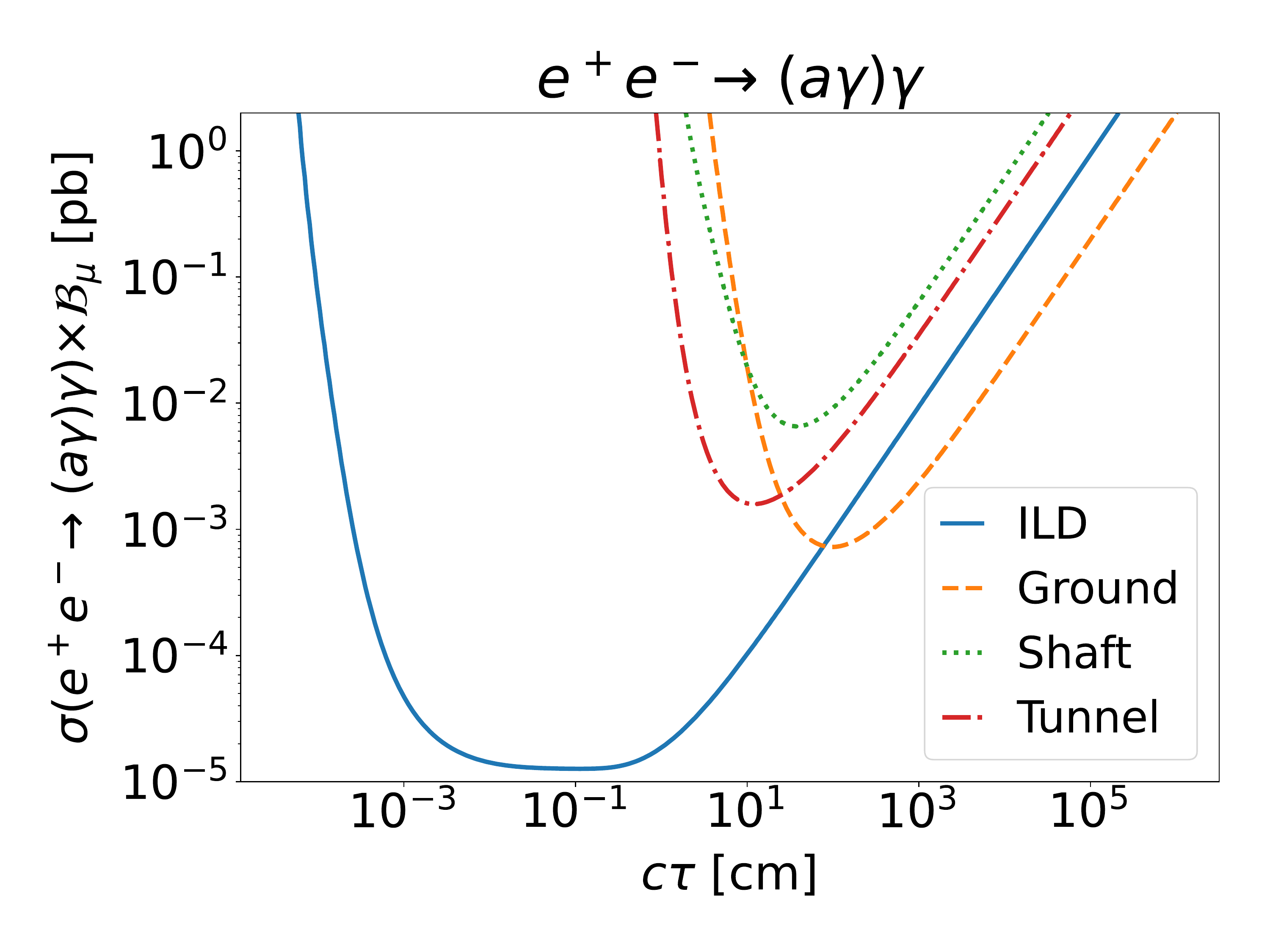}
    \caption{Contours of $N_a = 3$ ALPs with $m_a = 300\,$MeV decaying within various ILC detectors, as a function of the production cross section, $\sigma$, and the proper lifetime, $c\tau_a$. Shown are the production channels $e^+e^- \to a\gamma$ (left) and $e^+e^- \to Z\gamma \to (a\gamma)\gamma$ (right) at $\sqrt{s} = 250\,$GeV and with $\mathcal{L} = 250\,$fb$^{-1}$. Predictions are made for the ILD (blue, \sw{plain}) and far detectors placed in the Shaft (green, \sw{dotted}), in the Tunnel (red, \sw{dot-dashed}) and on the Ground (orange, \sw{dotted}). \sw{The branching ratio of the ALP into muons is indicated by $\mathcal{B}_{\mu}$.}
    \label{fig:sigma-ctau}}
\end{figure}

\paragraph{Generalization} For a less model-dependent interpretation of these results, in Fig.~\ref{fig:sigma-ctau} we show contours of $N_a = 3$ as a function of the production cross section $\sigma$ and the proper lifetime $c\tau_a$ of the ALP. 
 An ideal ILC experiment would probe the area above these contours. Since these results only depend on the production kinematics and the lifetime of the ALP, any (pseudo-)scalar particles with flavor-hierarchical couplings and masses well below the weak scale should lead to very similar results. This allows us to conclude more generally that only far detectors with a very large geometric acceptance $\Omega\cdot \langle r \rangle$ can improve the sensitivity to such LLPs, provided that the production cross section is large enough to probe particles with lifetimes beyond the reach of the ILD.
 
\paragraph{Experimental sensitivity} In our estimates, we have assumed that every ALP that decays inside the detector is reconstructed. In reality, however, the reconstruction of a displaced vertex of charged tracks from $a\to \ell^+\ell^-$ decays depends on the track length, angular separation and boost of the charged particles. The effective detection volume is thus smaller than the detector acceptance. This effect is particularly important for detectors that are relatively thin and far away from the production point, where the decay probability is proportional to the thickness, $\langle \mathds{P} \rangle \sim \langle r \rangle$.

\begin{figure}[t!]
    \centering
    \includegraphics[width=0.5\textwidth]{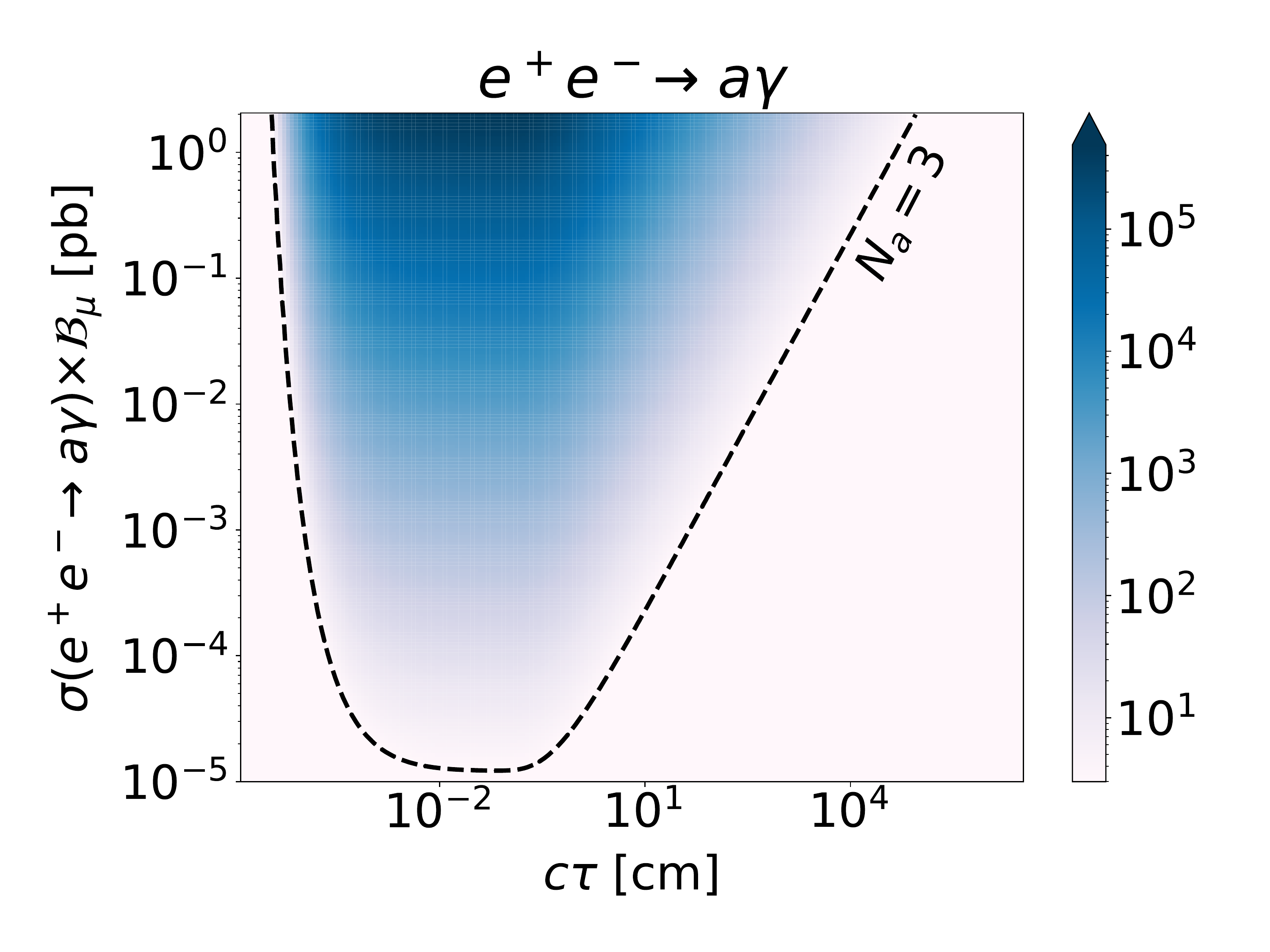}\hspace*{0.1cm}
    \includegraphics[width=0.5\textwidth]{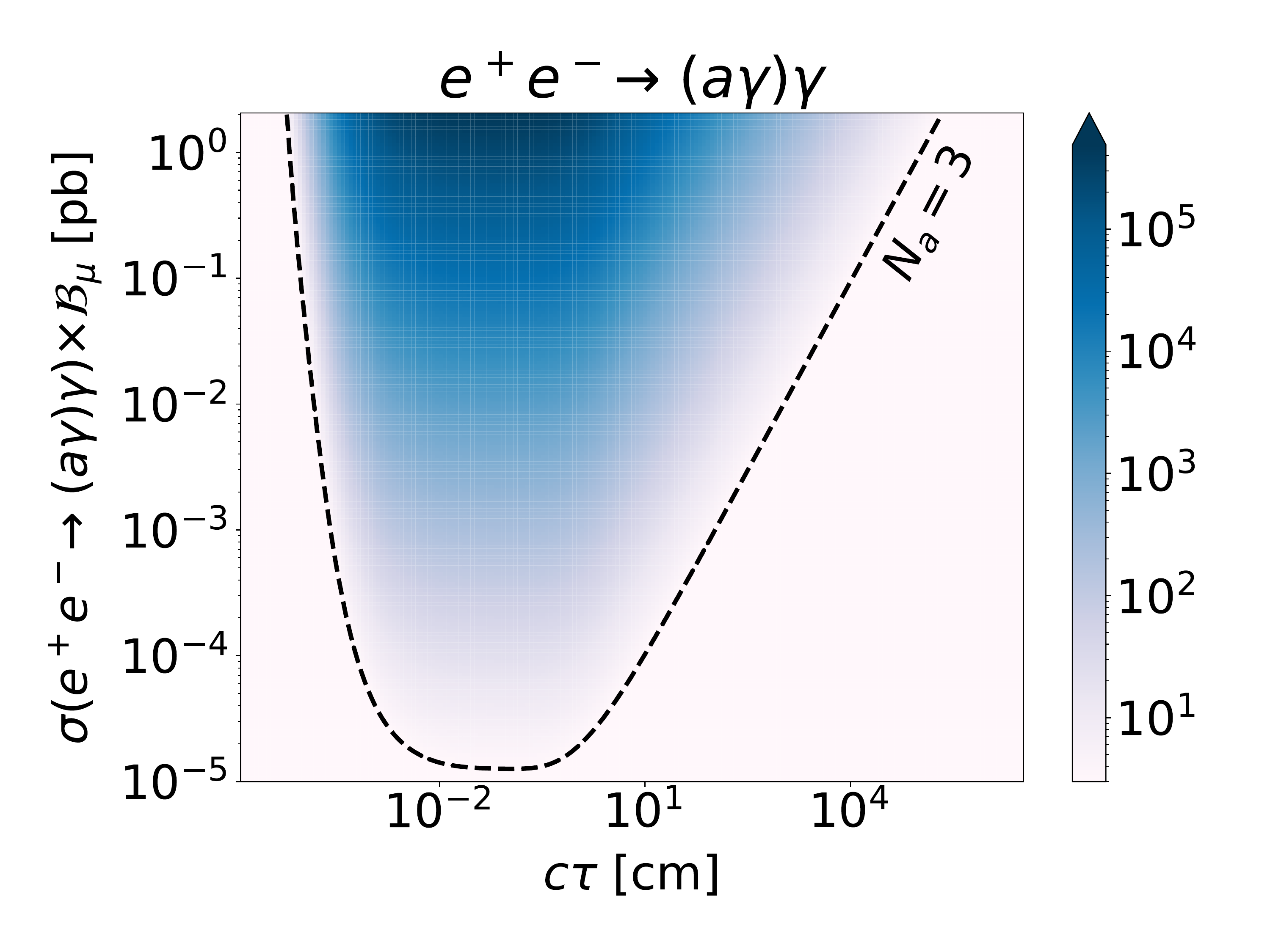}
    \caption{Event rates $N_a$ of ALPs with $m_a = 300\,$MeV decaying within the ILD, as a function of the production cross section, $\sigma$, and the proper lifetime, $c\tau_a$. Shown are the production channels $e^+e^- \to a\gamma$ (left) and $e^+e^- \to Z\gamma \to (a\gamma)\gamma$ (right) at $\sqrt{s} = 250\,$GeV and with $\mathcal{L} = 250\,$fb$^{-1}$.  \sw{The branching ratio of the ALP into muons is indicated by $\mathcal{B}_{\mu}$.}\label{fig:sigma-ctau-heat}}
\end{figure}

To illustrate the dependence of the sensitivity on the event rates, in Fig.~\ref{fig:sigma-ctau-heat} we show the predicted number of ALPs decaying within the ILD as a function of the production cross section and lifetime. The event rate decreases proportionally with the cross section and anti-proportionally with the lifetime in the limit of long decay lengths, as expected from~\eqref{eq:na} and~\eqref{eq:decay-prob-approx}.
 
The signal sensitivity is also expected to be reduced in the presence of background. At hadron colliders, the dominant source of background for displaced charged vertices are meson decays. At the ILC, mesons are produced at comparably lower, but still substantial rates, for example from hadronic $Z$ boson decays. Further background sources could be misreconstructed muon decays originating from $e^+ e^-$ collisions or from cosmic rays. Ideas to tame such and other backgrounds have been discussed for \belletwo in Ref.~\cite{Dreyer:2021aqd}. Notice that background can affect not only the absolute, but also the relative sensitivity of the various detectors. A dedicated background analysis goes beyond the scope of this work, but is crucial to determine the ultimate sensitivity of the ILC to long-lived particles.

\section{ILC versus Belle\,II}
\label{SEC:ilc-vs-belle2}
Long-lived particles with masses below a few GeV could well be observed at flavor experiments. ALPs produced through gauge interactions and decaying into leptons can be efficiently searched for in meson decays $B \to Ka, a\to \mu^+\mu^-$ or $K \to \pi a, a\to \mu^+\mu^-$. In these decays, the ALP can be reconstructed from a prompt or displaced vertex, depending on its lifetime. At $e^+ e^-$ experiments, ALPs could also be produced directly via $e^+ e^- \to \gamma a, a\to \ell^+\ell^-$, resulting in a signature of a displaced vertex and a prompt photon. This signature has recently been explored for a similar scenario with dark photons at \belletwo and found to probe very small couplings~\cite{Ferber:2022ewf}. An interpretation for ALPs would be a worthwhile endeavor, but goes beyond the scope of this work.

\paragraph{ALPs in meson decays} The currently strongest bounds on long-lived ALPs in $B \to Ka, a\to \mu^+\mu^-$ have been obtained by the LHCb collaboration~\cite{LHCb:2016awg}. The \belletwo experiment can search for these decays in $e^+ e^-$ collisions at the $\Upsilon(4S)$ resonance $\sqrt{s} = 10.58\,$GeV and potentially extend the reach for long-lived light pseudo-scalars~\cite{Filimonova:2019tuy,Kachanovich:2020yhi,Ferber:2022rsf}. It is thus interesting to compare the projected sensitivity for \belletwo with our predictions for the ILC. While both experiments rely on electron-positron collisions, they probe ALPs in different production channels and phase-space regions.

\begin{figure}[t!]
    \centering
   \raisebox{1.7cm}{ \includegraphics[width=0.44\textwidth]{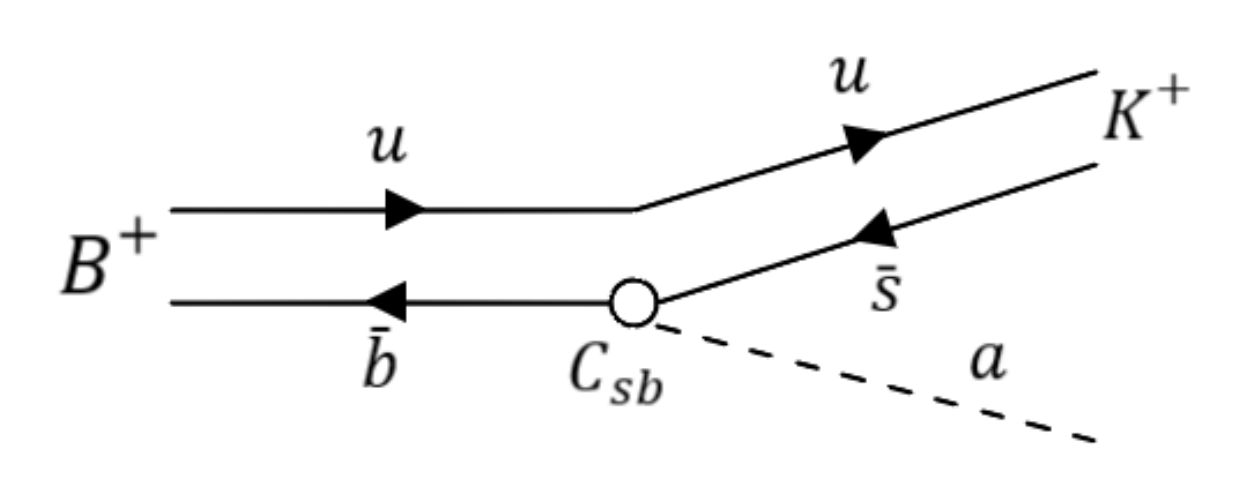}}\hspace*{1.3cm}
         \includegraphics[width=0.36\textwidth]{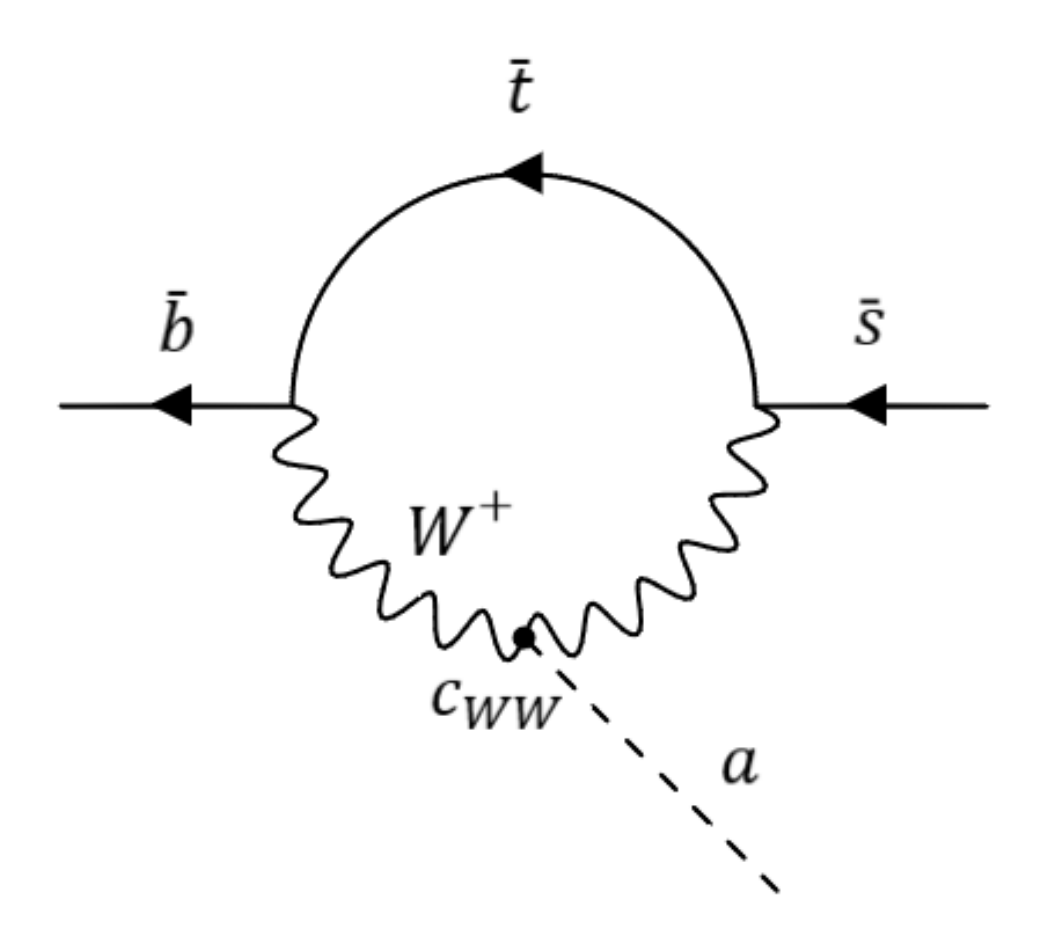}    
    \caption{Feynman diagrams for ALP production in meson decays $B^+ \to K^+ a$ (left) and loop-induced flavor-changing coupling $C_{sb}$ of an ALP with gauge coupling $c_{WW}$ (right). }
    \label{fig:feynman-belle2}
\end{figure}

For ALPs with coupling $c_{WW}$, the decay $B\to K a$ is induced by flavor-changing neutral currents at one-loop level, as illustrated in Fig.~\ref{fig:feynman-belle2}.
 At energies $\mu < \mu_w$ below the weak scale, the ALP coupling to left-handed bottom and strange quarks is described by a local operator in the effective Lagrangian
\begin{align}
    \mathcal{L}_{\rm eff}(\mu < \mu_w) \supset C_{sb}(\mu)\,\frac{\partial^{\mu} a}{f_a} (\bar{s}_L \gamma_{\mu} b_L) + h.c.
\end{align}
At the weak scale $\mu_w \approx m_W$, the Wilson coefficient reads
\begin{align}\label{eq:csb-coupling}
C_{sb}(\mu_w) = - V_{ts}^\ast V_{tb}\, \frac{\alpha_t}{4\pi} \frac{3\alpha}{2\pi s_w^2}\frac{1 - x_t + x_t \ln x_t}{(1-x_t)^2}\,c_{WW}(\mu_w)\,,
\end{align}
where $\alpha_t = y_t^2/4\pi$ and $x_t = m_t^2/m_W^2$. The evolution down to the bottom mass scale $m_b$ can be neglected, so that $C_{sb}(\mu_w) \approx C_{sb}(m_b)$~\cite{Bauer:2020jbp}. The decay rate is finally given by 
\begin{align}\label{eq:b-to-ka}
        \Gamma(B\to K a) &= \frac{m_B}{16\pi}\left(\frac{C_{sb}(m_b)}{f_a}\right)^2 f_0^2\left(m_a^2\right)\left(1-\frac{m_K^2}{m_B^2}\right)^2\lambda^{1/2}(m_B^2,m_K^2,m_a^2)\,,
    \end{align}
with the kinematic function $\lambda(a,b,c) =a^2+b^2+c^2 -2(ab + ac + bc)$ and the $B\to K$ scalar form factor $f_0(q^2)$ at momentum transfer $q^2 = m_a^2$~\cite{Gubernari:2018wyi}.

\paragraph{\belletwo's sensitivity to long-lived ALPs} To determine the expected event rates for $B^+\to K^+ a, a\to \mu^+\mu^-$, we closely follow the analysis of Ref.~\cite{Dreyer:2021aqd}. Based on a sample of 10,000 simulated events,\footnote{We thank Torben Ferber for sharing his event samples with us.} we determine the number of ALPs that decay within the \belletwo tracking detector as
\begin{align}\label{eq:na-belle2}
N_a = N_{B\bar{B}} \,\mathcal{B}\left(B\to K a\right)\mathcal{B}\left(a \to \mu^+\mu^-\right)\langle \mathds{P} \rangle.
\end{align}
Here $N_{B\bar{B}} = 5\cdot 10^{10}$ is the number of $B\bar{B}$ pairs expected at \belletwo for an integrated luminosity of $50\,$ab$^{-1}$ and $\langle \mathds{P} \rangle$ is the average probability for an ALP to decay within the \belletwo detector volume. We consider only the tracking parts of the \belletwo detector, i.\,e., the PXD, SVD, and CDC, which feature a good reconstruction efficiency for displaced vertices~\cite{Belle-II:2018jsg}. Furthermore, we reduce the outer radius of the CDC from $113\,$cm to $60\,$cm, in order to allow for long enough tracks that can be reconstructed with less detector resolution than in the PXD and SVD.\footnote{We have not made such a request for the ILC detectors, which leads to a slight bias in the comparison with \belletwo.} The resulting effective detection volume is defined as
\begin{itemize}
\item \belletwo:$\quad z \in [-55,140]\,$cm,$\quad \rho \in [0.9,60]\,$cm,$\quad \theta \in [17,150]\degree\,$,
\end{itemize}
with the electron beam pointing along the positive $z$ direction and the $e^+e^-$ collision point placed at $(z,\rho) = (0,0)$.

\begin{figure}[t!]
    \centering
    \includegraphics[width=0.55\textwidth]{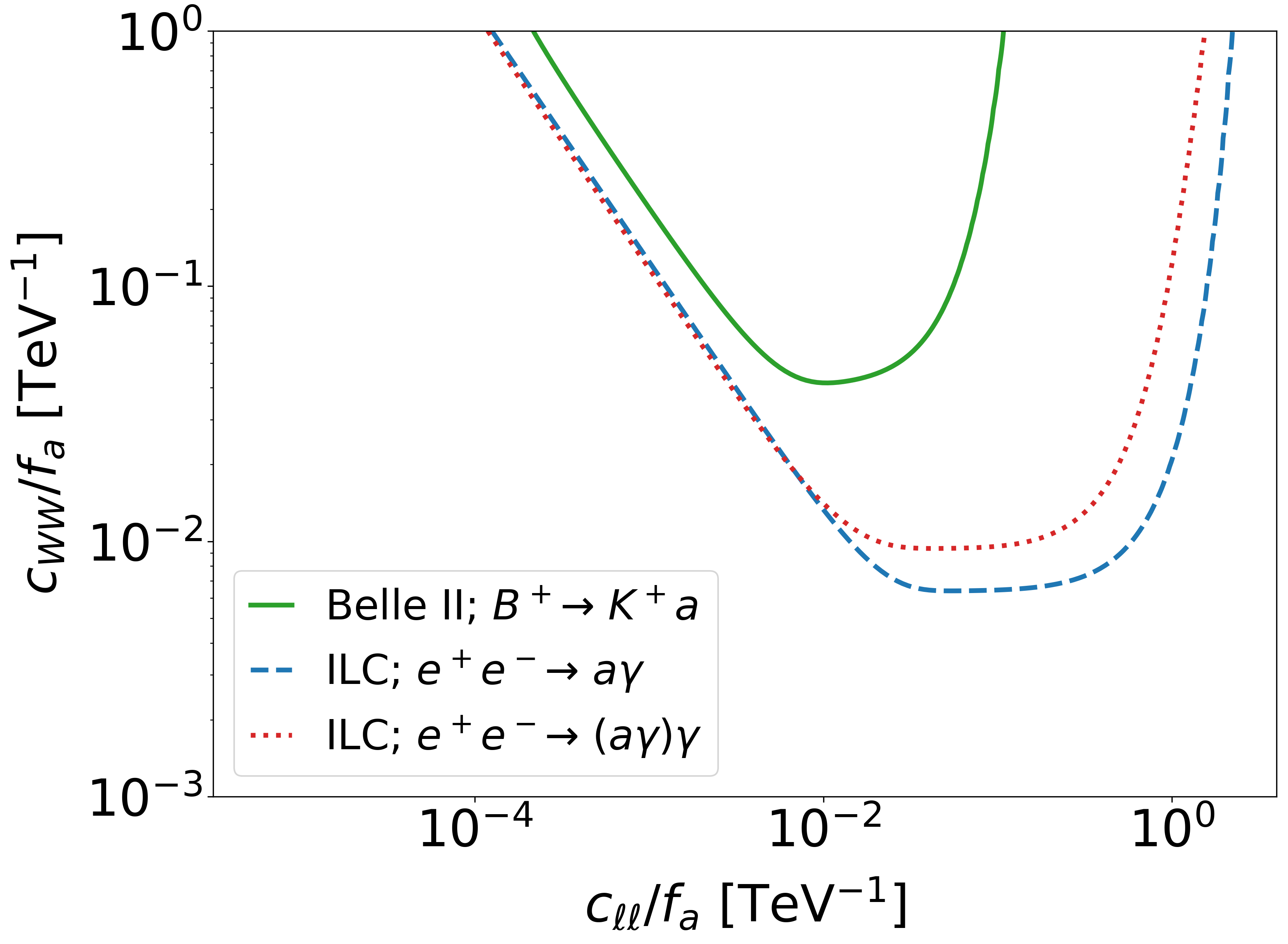}
    \caption{Sensitivity projections for long-lived ALPs with $m_a = 300\,$MeV at the ILC and at \belletwo, as a function of the couplings $c_{WW}/f_a$ and $c_{\ell\ell}/f_a$. Shown are $N_a = 3$ contours for ALP decays within the ILD for $e^+e^- \to a\gamma$ (blue, \sw{dashed}) and $e^+e^- \to Z\gamma \to (a\gamma)\gamma$ (red, \sw{dotted}) production at $\sqrt{s} = 250\,$GeV and with $\mathcal{L} = 250\,$fb$^{-1}$. For comparison, $N_a = 3$ contours are shown for $B^+\to K^+ a,\, a\to \mu^+\mu^-$ decays within the tracking chambers of the \belletwo detector with $\mathcal{L} = 50\,\text{ab}^{-1}$ \sw{(green)}. \label{fig:ilc-vs-belle2}}
\end{figure}

In Fig.~\ref{fig:ilc-vs-belle2}, we show a contour of $N_a = 3$ ALP decays as expected from $B^+\to K^+ a, a\to \mu^+\mu^-$ decays at \belletwo for $50\,$ab$^{-1}$ of data.
 The area above the contour can be probed with meson decays. The results are shown for an ALP mass $m_a = 300\,$MeV and as a function of the ALP couplings $c_{WW}$ and $c_{\ell\ell}$. From~\eqref{eq:csb-coupling} and \eqref{eq:b-to-ka}, it is apparent that the production rate of the ALPs scales as $\mathcal{B}(B\to K a) \sim (c_{WW}/f_a)^2$. \sw{The event rate $N_a$ also depends on the lepton coupling through the branching ratio into muons and through the lifetime of the ALP, see~\eqref{eq:na-belle2}. For large decay lengths or small couplings, see \eqref{eq:decay-prob-approx}, the event rate scales as $N_a \sim (c_{WW}c_{\ell\ell}/f_a^2)^2$, resulting in a straight line on the double-logarithmic scale. Notice that in the long-lifetime limit the event rate only depends on the partial width into muons, $\Gamma(a\to \mu^+\mu^-)$, not on the branching ratio, $\mathcal{B}(a\to \mu^+\mu^-) = \Gamma(a\to \mu^+\mu^-)/\Gamma_a$. The dependence on the total width, $\Gamma_a$, cancels with the decay probability, $\langle \mathds{P} \rangle \sim \Gamma_a$.}

\paragraph{Sensitivity at \belletwo versus ILC} For comparison, in Fig.~\ref{fig:ilc-vs-belle2} we also show the expected reach of the ILD for ALPs produced via $e^+e^- \to a\gamma$ and $e^+e^- \to Z\gamma \to (a\gamma)\gamma$. For decay lengths $\langle d \rangle \gg 2\,$m, the event rate \eqref{eq:na} scales with the couplings as $N_a \sim (c_{WW}c_{\ell\ell}/f_a^2)^2$, as for meson decays. In our benchmark scenario with $m_a = 300\,$MeV and $c_{\ell\ell}/f_a \ll 1/$TeV, we find $N_a = 3$ events for 
\begin{align}
& \text{\belletwo}\ B^+ \to K^+ a:&&|c_{WW}c_{\ell\ell}|/f_a^2 = 2.0\cdot 10^{-4}/\text{TeV}^2\\\nonumber
& \text{ILD }e^+ e^- \to a\gamma:&&|c_{WW}c_{\ell\ell}|/f_a^2 = 1.24\cdot 10^{-4}/\text{TeV}^2\\\nonumber
& \text{ILD }e^+ e^- \to (a\gamma)\gamma:&& |c_{WW}c_{\ell\ell}|/f_a^2 = 1.17\cdot 10^{-4}/\text{TeV}^2\,,
\end{align}
assuming luminosities of $50\,$ab$^{-1}$ for \belletwo and $250\,$fb$^{-1}$ for the ILC. In Fig.~\ref{fig:ilc-vs-belle2}, these values define the straight lines at small $c_{\ell\ell}/f_a$. At the ILD, the expected sensitivity to small couplings is thus improved by about $60\%$ compared with \belletwo. More generally, the ILD will be more sensitive to long-lived ALPs than \belletwo throughout the parameter space. The gain is particularly large at small $c_{WW}$, due to the larger production rates and the excellent detector coverage around the ILC interaction point. For $e^+e^- \to a\gamma$, the sensitivity at small $c_{WW}$ is larger than for $e^+e^- \to Z\gamma \to (a\gamma)\gamma$, due to the larger cross section~\eqref{eq:prod-xs}.  

From the positions of the minima along the contours, we deduce that the ILD is most sensitive at larger $c_{\ell\ell}$ or shorter lifetimes compared to \belletwo, due to the higher boost of the ALPs at the ILC. At very short lifetimes, both \belletwo and the ILC can efficiently search for promptly decaying particles and complement the reach of displaced vertex searches.

\paragraph{Sensitivity at other current and future experiments} \sw{Besides $e^+e^-$ colliders, ALPs can be searched for in other environments like hadron colliders, fixed-target experiments, or in astrophysics observables. For ALPs with MeV-GeV masses and pure couplings to gauge bosons or leptons, bounds have been derived for instance in Refs.~\cite{Bauer:2017ris,Dolan:2017osp,Gori:2020xvq,Bauer:2021mvw,Ferber:2022rsf}. Some of them constrain parts of the $\{c_{\ell\ell},c_{WW}\}$ parameter space, mostly in the upper right corner of Fig.~\ref{fig:ilc-vs-belle2}. Fixed-target experiments probe the smallest $c_{WW}$ couplings. However, an extra $c_{\ell\ell}$ coupling shortens the ALP's lifetime, such that it decays before the detector and is vetoed at most long-baseline experiments. We therefore do not expect additional stringent bounds on the $\{c_{\ell\ell},c_{WW}\}$ space from such experiments. Deriving the exact bounds would require a dedicated re-analysis of each search, because the production rate, decay modes and lifetime of the ALP all change compared to the case of pure lepton or gauge couplings. As our main goal is to explore the potential added value of far detectors compared to the main detector at the ILC, we leave the detailed comparison with other experiments for future research.}

\sw{Future experiments and annexes to existing experiments can extend the current sensitivity to ALPs. At the FCC-ee, which is closest in setup to the ILC, the projected sensitivity to long-lived ALPs~\cite{Bauer:2018uxu,Alimena:2022hfr} is comparable with that of the ILC main detector. Searches for ALP decays to photons can further enhance the sensitivity to the ALP coupling $c_{WW}$ at both experiments. As for the ILC, far detectors at Belle~II cannot provide much gain over the main Belle~II detector, which has an excellent angular coverage and detection efficiency~\cite{Dreyer:2021aqd}. On the other hand, proposed far detectors around the LHC such as FASER, MATHUSLA, CODEX-b or Anubis, as well as future searches at fixed-target experiments have a high potential to probe feebly coupling ALPs~\cite{Agrawal:2021dbo}. Analy\-zing the complementarity of these searches is a community effort, which already shows that much of the currently unexplored parameter space could be tested in the near future.}

\section{Conclusions}
\label{SEC:conclusions}
In this work we have \sw{been guided by} the question whether near or far detectors are more sensitive to long-lived particles produced at future electron-positron colliders. \sw{To explore this question concretely, we} have performed a comparative study for sub-GeV axion-like particles produced via $e^+e^- \to a\gamma$ and $e^+e^- \to Z\gamma \to (a\gamma)\gamma$ at the ILC. The two production channels lead to different kinematic distributions: ALPs from $e^+e^- \to a\gamma$ are emitted mostly \sw{in the central region}, while ALPs from $Z$ boson decays in $e^+e^- \to Z\gamma \to (a\gamma)\gamma$ are boosted along the beam axis. These kinematics affect the relative sensitivity of detectors placed in the forward direction or in the central region. We have demonstrated this effect for two realistic far detector options, one placed in a supply tunnel in the forward region and one placed in a vertical shaft above the ILC interaction point.

Neither of these underground far detectors can enhance the sensitivity to \sw{ALPs} compared to the main detector ILD. This is mostly due to the almost full angular coverage and relative thickness of the ILD, which captures a large number of particles with decay lengths \sw{even} beyond its geometric extensions. To enhance the sensitivity, any far detector must have a large angular coverage and a larger radial thickness than the ILD. For an underground experiment, such conditions can realistically only be fulfilled on the ground. \sw{Even with a (technically unrealistic)} kilometer-sized cuboid on the surface above the ILC detector hall, the sensitivity to long\sw{-lived ALPs} increases by \sw{at most} a factor of four. Moderate improvements with far detectors thus require a substantial extra construction effort.

A key result of our analysis is that the ILC main detector itself is very sensitive to long-lived particles. For light \sw{ALPs} produced at picobarn rates, the ILD can probe decay lengths up to $100\,$m with a luminosity of $250\,$fb$^{-1}$. For smaller cross sections in the femtobarn range, the sensitivity still reaches decay lengths of about $10\,$cm.

To quantify the gain of a high-energy lepton collider over current low-energy experiments, we have compared our predictions for the ILC with searches for long-lived ALPs produced from meson decays at \belletwo. At long lifetimes, the ILC can improve the sensitivity to the product of ALP couplings, $c_{WW} c_{\ell\ell}/f_a^2$, by about $60\%$ compared to \belletwo. At shorter lifetimes, where the ILC reaches its maximum sensitivity, the ILD can probe ALPs produced through couplings $c_{WW}$ that are an order of magnitude smaller than the reach of \belletwo with its total expected luminosity of $50\,$ab$^{-1}$.

These results apply more generally for ALPs that are effectively massless at the ILC, \emph{i.\,e.}, ALPs with masses below a few GeV, provided that the lifetime is rescaled accordingly. Heavier ALPs are produced with tendencially lower boosts and have shorter lifetimes. Both features lead to enhanced event rates in near detectors. \sw{The main results of our analysis rely mostly on the detector geometry and position, rather than on the exact production mode and kinematic distribution of the ALP. Therefore we expect similar relative sensitivities of the main and far detectors to other LLPs with comparable lifetimes.}

All our statements are based on pure event rates, assuming a perfect detector with zero background and excellent reconstruction efficiency. Realistic predictions of the ultimate sensitivity require a dedicated analysis of background and detector effects, which goes beyond the scope of this study. However, our predictions for the ILC under ideal conditions allow for a better comparison with other electron-positron colliders. For similar collision energies, the main messages from this work should apply as well to the FCC-ee and CEPC. They can serve as a guideline to optimize the LLP program at the next high-energy collider.

\acknowledgments
We thank Hitoshi Murayama and Maxim Perelstein for their enthusiasm for this study and for interesting discussions. We also thank Torben Ferber for providing us with event si\-mu\-lations for \belletwo and for helpful discussions. RS acknowledges support of the \emph{Deutsche Forschungsgemeinschaft} (DFG) through the research training group \emph{Particle Physics Beyond the Standard Model} (GRK 1940). The research of SW is supported by the DFG under grant no. 396021762–TRR 257.



\bibliographystyle{JHEP_improved}
\bibliography{main}

\end{document}